\documentclass[11pt]{article}
\usepackage{amsmath,amsthm,amssymb,cite,enumerate}
\usepackage[a4paper,left=0.5in,right=1in]{geometry}
\usepackage[usenames]{color}
\usepackage{graphicx}
\usepackage{epsfig}
\usepackage{dcolumn}% Align table columns on decimal point
\usepackage{bm}% bold math
\usepackage{authblk} %for authors with multiple affiliations
\usepackage{soul} % enables one to strike out text 

\usepackage{t1enc}	%for GHP symbols

\begin{document}

\def \H {\phi}
\def \y {y}
\def \d {{\rm d}}
\def \ff {\psi}

\def \TH {{\tr{the }}}

\def \bm {\mbox{\boldmath{$m$}}}
\def \bmi #1 {\mbox{\boldmath{$m^{({#1})}$}}}

\def \bF {\mbox{\boldmath{$F$}}}
\def \bA {\mbox{\boldmath{$A$}}}
\def \cF {\mbox{\boldmath{$\cal F$}}}
\def \bH {\mbox{\boldmath{$H$}}}
\def \bC {\mbox{\boldmath{$C$}}}
\def \bSS {\mbox{\boldmath{$S$}}}
\def \bS {\mbox{\boldmath{${\cal S}$}}}
\def \bV {\mbox{\boldmath{$V$}}}
\def \bff {\mbox{\boldmath{$f$}}}
\def \bT {\mbox{\boldmath{$T$}}}
\def \bk {\mbox{\boldmath{$k$}}}
\def \bl {\mbox{\boldmath{$\ell$}}}
\def \bn {\mbox{\boldmath{$n$}}}
\def \bbm {\mbox{\boldmath{$m$}}}
\def \tbbm {\mbox{\boldmath{$\bar m$}}}
\def \bet {\mbox{\boldmath{$\eta$}}}
\def \bnt {\mbox{\boldmath{$\tilde n$}}}

\def \bmd #1 {\mbox{\boldmath{$m_{(#1)}$}}}

\def \T {\bigtriangleup}
\newcommand{\msub}[2]{m^{(#1)}_{#2}}
\newcommand{\msup}[2]{m_{(#1)}^{#2}}

\def \tbl {\mbox{\boldmath{$\tilde \ell$}}}
\def \hbl {\mbox{\boldmath{$\hat \ell$}}}
\def \hbn {\mbox{\boldmath{$\hat n$}}}
\def \hbm #1 {\mbox{\boldmath{$\hat m^{({#1})}$}}}
\def \hL {\hat{L}}
\def \hN {\hat{N}}

\newcommand{\hM}[3] {{\stackrel{#1}{\hat{M}}}_{{#2}{#3}}}

\newcommand{\be}{\begin{equation}}
\newcommand{\ee}{\end{equation}}
\newcommand{\bea}{\begin{eqnarray}}
\newcommand{\eea}{\end{eqnarray}}

\newcommand{\beqn}{\begin{eqnarray}}
\newcommand{\eeqn}{\end{eqnarray}}
\newcommand{\AdS}{anti--de~Sitter }
\newcommand{\AAdS}{\mbox{(anti--)}de~Sitter }
\newcommand{\AAN}{\mbox{(anti--)}Nariai }
\newcommand{\AS}{Aichelburg-Sexl }
\newcommand{\pa}{\partial}
\newcommand{\pp}{{\it pp\,}-}
\newcommand{\ba}{\begin{array}}
\newcommand{\ea}{\end{array}}

\newcommand*\bR{\ensuremath{\boldsymbol{R}}}

\newcommand*\BF{\ensuremath{\boldsymbol{F}}}
\newcommand*\BR{\ensuremath{\boldsymbol{R}}}
\newcommand*\BS{\ensuremath{\boldsymbol{S}}}
\newcommand*\BC{\ensuremath{\boldsymbol{C}}}
\newcommand*\bg{\ensuremath{\boldsymbol{g}}}
\newcommand*\bE{\ensuremath{\boldsymbol{E}}}

\newcommand*\bh{\ensuremath{\boldsymbol{h}}}
\newcommand*\bZ{\ensuremath{\boldsymbol{Z}}}

\newcommand{\M}[3] {{\stackrel{#1}{M}}_{{#2}{#3}}}
\newcommand{\Mt}[3] {{\stackrel{#1}{\tilde M}}_{{#2}{#3}}}
\newcommand{\m}[3] {{\stackrel{\hspace{.3cm}#1}{m}}_{\!{#2}{#3}}\,}

\newcommand{\MM}[2] {{\stackrel{#1}{M}}_{#2}}

%older version for Ms

\newcommand{\ovM}[1]{\overset{\,#1}{M}}
\newcommand{\Mi}{\ovM{i}}
\newcommand{\Mj}{\ovM{j}}
\newcommand{\Mk}{\ovM{k}}

\newcommand{\tr}{\textcolor{red}}
\newcommand{\tb}{\textcolor{blue}}
\newcommand{\tg}{\textcolor{green}}

\newcommand{\thorn}{\mathop{\hbox{\rm \th}}\nolimits}

\def\a{\alpha}
\def\g{\gamma}
\def\de{\delta}

\def\b{{\kappa_0}}

\def\E{{\cal E}}
\def\B{{\cal B}}
\def\R{{\cal R}}
\def\F{{\cal F}}
\def\L{{\cal L}}

\def\e{e}
\def\bb{b}

\def\tu{\tilde u}
\def \tH {\tilde H}
\def \tL {\tilde L}
\def \t {\tilde}

\newtheorem{theorem}{Theorem}[section] 
\newtheorem{cor}[theorem]{Corollary} 
\newtheorem{lemma}[theorem]{Lemma} 
\newtheorem{proposition}[theorem]{Proposition}
\newtheorem{definition}[theorem]{Definition}
\newtheorem{remark}[theorem]{Remark}

\title{Kerr-Schild double copy for Kundt spacetimes of any dimension}

\author{Marcello Ortaggio\thanks{ortaggio (at) math (dot) cas (dot) cz}, Vojt\v ech Pravda\thanks{pravda@math.cas.cz}, Alena Pravdov\'a\thanks{pravdova@math.cas.cz} \\
	Institute of Mathematics, Czech Academy of Sciences \\ \v Zitn\' a 25, 115 67 Prague 1, Czech Republic}

\maketitle

\abstract{ 
We show that vacuum type N Kundt spacetimes in an arbitrary dimension admit a Kerr-Schild (KS) double copy. This is mostly done in a coordinate-independent way using the higher-dimensional Newman-Penrose formalism. We also discuss two kinds of non-uniqueness of an electromagnetic field corresponding to a given KS metric (i.e., its single copy) -- these originate, respectively, from the rescaling freedom in the KS vector and from the non-uniqueness of the splitting of the KS metric in the flat part and the KS part. In connection to this, we show that the subset of KS \pp waves admits both null and non-null electromagnetic single copies. Since vacuum type N Kundt spacetimes are universal solutions of virtually any higher-order gravities and null fields in such backgrounds are immune to higher-order electromagnetic corrections, the KS-Kundt double copy demonstrated in the present paper also applies to large classes of modified theories.
 }

\vspace{.2cm}
\noindent

%PACS 04.50.+h, 04.20.Jb, 04.40.Nr
%
%% 04.20.Jb  Exact solutions
%% 04.50.+h  in more than four dimensions
%% 04.40.Nr  Einstein-Maxwell spacetimes, spacetimes with fluids, radiation or classical fields

\tableofcontents

\section{Introduction}

\label{intro}
Recently, there has been a great interest in the notion of double copy, an approach in which gravity can be understood as two copies of a gauge theory. This approach originated from relating perturbative scattering amplitudes in a non-abelian gauge theory and gravity \cite{Bern08,Bern10}. 
Subsequently, it has been extended to relate exact vacuum solutions of Einstein gravity with solutions to the Maxwell equations in a flat spacetime \cite{MonOCoWhi14}.

The essential assumption for the original formulation \cite{MonOCoWhi14} of the double copy for classical GR is that an $n$-dimensional spacetime metric can be expressed in the so-called Kerr-Schild (KS) form
\be
\d s^2=g_{ab}\d x^a\d x^b=\d s^2_{\mbox{\tiny flat}}- 2 \H \bl\otimes\bl ,  
\label{KS}
\ee
where the ``background'' metric $\d s^2_{\mbox{\tiny flat}}=\eta_{ab}\d x^a\d x^b$ is flat (although not necessarily expressed in the Minkowskian coordinates), the KS covector field $\bl$ is null (with respect to both metrics $\bet$ and $\bg$), and $a, b\ldots=0,\ldots,n-1$.

In vacuum (i.e., in the Ricci-flat case, considering the Einstein equations), several results for KS metrics in arbitrary dimension were obtained in \cite{OrtPraPra09}. In particular, the null congruence defined by $\bl$ must be {\em geodesic}, which will thus be understood from now on.
The KS ansatz \eqref{KS} then restricts possible algebraic types of KS spacetimes.
In fact in arbitrary dimension, Ricci-flat KS metrics are Weyl-type\footnote{The Weyl type is a generalization of the Petrov type to higher dimensions in the algebraic classification of \cite{Coleyetal04} (see also \cite{OrtPraPra13rev} for a review).} D or II when the KS vector $\bl$ is expanding. When $\bl$ is non-expanding, the KS congruence becomes Kundt, and the Weyl type is~N (or O for a flat spacetime).

Thus clearly, the KS ansatz is compatible only with a subset of solutions to the Einstein equations. However, it includes several spacetimes of great physical importance, such as  Schwarzschild and Kerr black holes, their higher-dimensional counterparts  \cite{Tangherlini63,MyePer86} and type~N  \pp-waves.
Note that while in four dimensions all Ricci-flat KS spacetimes are known \cite{KerSch652,DebKerSch69,Urbantke72,Debney73,Debney74},\footnote{Some errors in \cite{Debney73} are amended in \cite{Debney74}.} it is not so in higher dimensions.

It has been shown \cite{MonOCoWhi14} that for {\em stationary} KS spacetimes, using an appropriate normalization of the KS covector field $\bl$, the vacuum Einstein equations for the metric \eqref{KS} imply\footnote{Note, however, that the Maxwell equations themselves do not guarantee that metric~\eqref{KS} obeys the vacuum Einstein equations, see \cite{MonOCoWhi14}.} that the Maxwell equations for an electromagnetic field 
\be
 \bA = \H \bl ,
 \label{A_dc}
\ee
hold in the flat background spacetime $\d s^2_{\mbox{\tiny flat}}$. Thus the electromagnetic field is constructed by multiplying the scalar $\H$ by a single copy of the covector $\bl$, while the gravitational field (i.e., the ``perturbation'' $-2 \H \bl\otimes\bl$ of the flat metric $
\d s^2_{\mbox{\tiny flat}}$ in~\eqref{KS}) is obtained by  multiplying the scalar $\H$ by two copies of $\bl$. Recently, it has been pointed out \cite{Eassonetal23} (see also \cite{OrtSri23}) that
in fact the KS double copy of stationary, expanding spacetimes can be traced back to the well-known relation between test Maxwell fields and Killing vector fields \cite{Papapetrou66,Wald74} (at least in four dimensions this applies, more generally, to all expanding KS metrics \cite{Eassonetal23}, since they necessarily possess at least one Killing vector field \cite{DebKerSch69} -- not necessarily timelike).
It is thus interesting to study the KS double copy also in time-dependent spacetimes,\footnote{More precisely, in settings such that the potential~\eqref{A} is not gauge-equivalent to a Killing covector.} for which the method of \cite{Papapetrou66,Wald74} cannot in general be applied. Although the KS double copy has been extended also to certain time-dependent solutions such as \pp waves \cite{MonOCoWhi14} (see also \cite{BahStaWhi20}), works devoted to the KS double copy have focused mostly on stationary Weyl type D KS spacetimes \cite{MonOCoWhi14,BahLunWhi17}, since these are of particular relevance in the context of black holes (see also \cite{Eassonetal23} for the discussion of an example of a type II KS spacetime in the context of the double copy). {On the other hand, from the viewpoint of scattering amplitudes \cite{Bern08,Bern10}, one would be more interested in the role of gravitons and thus in gravitational waves.}

In this paper, we will focus on the detailed analysis of the only remaining algebraic type compatible with the KS ansatz~\eqref{KS} -- namely, we will study  Weyl type N spacetimes of arbitrary dimension in the context of KS double copy.

As mentioned above, results of \cite{OrtPraPra09} imply that Ricci-flat KS type N spacetimes are necessarily Kundt. By definition, Kundt spacetimes admit a null geodesic vector field with vanishing optical scalars shear, twist, and expansion.\footnote{For Ricci-flat KS spacetimes, vanishing expansion implies that shear and twist vanish as well and it is thus a sufficient condition for a spacetime being Kundt \cite{OrtPraPra09}.}  In fact, \cite{OrtPraPra09} also shows that all type N Ricci-flat Kundt spacetimes are compatible with the KS ansatz~\eqref{KS}. Thus in arbitrary dimension for Ricci-flat type N spacetimes, being Kundt and admitting KS metric~\eqref{KS}  are equivalent conditions. 
In contrast with the expanding stationary case, for Kundt spacetimes, the KS double copy cannot be in general traced to electromagnetic test fields constructed from Killing vectors and thus in this case the double copy is a genuinely distinct procedure.\footnote{For example, generic \pp waves admit only one Killing vector $\bl$, which is covariantly constant and thus cannot be used to produce a non-trivial test Maxwell field with the method of \cite{Papapetrou66,Wald74} (i.e., $\bF=\d A$ vanishes if $\bA=\bl$).}

Kundt metrics \cite{Kundt61,Kundt62} have played an important role as exact solutions describing, in particular, plane \cite{BPR} and \pp waves \cite{Brinkmann25}, as well as the gravitational field produced by light-like sources \cite{Peres60,Bonnor69,AicSex71}. Higher-dimensional extensions have been studied, e.g., in \cite{Coleyetal03,ColHerPel06,PodZof09} (cf. also \cite{OrtPraPra13rev} and references therein).
Type N Ricci-flat Kundt spacetimes consist of two disjoint classes of gravitational waves -- Kundt waves and  \pp-waves. Denoting by $\bl$ the Kundt vector field, \pp-waves can be characterized by the property $\ell_{a;b}=\ell_a p_{,b}$, where $p$ is a scalar function ($\bl$ is determined up to a rescaling, which can be used to set $p=0$), while for Kundt waves one has $\ell_{a;b}=\ell_a p_b+q_a\ell_b$, where $p_a$ and $q_a$ are covector fields with $q_a\ell^a=0\neq q_a$ (in this case, the scaling freedom can be used to set, e.g., $p_a=q_a$ or $p_a=0$, if desired) -- cf. also section~\ref{subsec_double} and \cite{OrtPraPra13rev}.

In this paper, we establish that the double copy holds for all type N Ricci-flat Kundt spacetimes in an arbitrary dimension. In order not to be limited by a specific choice of coordinates for Kundt spacetimes,  first we approach the problem and express, e.g., the double-copy compatibility conditions using the higher-dimensional Newman-Penrose  (NP) formalism \cite{Pravdaetal04,Coleyetal04vsi,OrtPraPra07}, resorting to coordinates only at a later stage. 

We also study the non-uniqueness of the single-copy electromagnetic field corresponding to a given KS metric. Two distinct reasons for this non-uniqueness are: i) the possibility of rescaling the KS vector $\bl$ (with a simultaneous appropriate rescaling of the KS function $\H$); ii) the non-uniqueness of the splitting of the KS metric in the flat part and the KS part of the metric. The first non-uniqueness i) leads to the infinite non-uniqueness of an electromagnetic field arising as a single copy from a given KS Kundt metric. As a consequence of ii), a single copy of a \pp wave can be either a null or a non-null electromagnetic field.

The KS double copy has been considered also in the context of nonlinear electrodynamics, especially as a tool for constructing regular black holes \cite{PasTse20,EasKeeMan20,MkrSva22}. It is therefore interesting to point out that the KS-Kundt double copy demonstrated in the present paper also applies to higher-order theories (except for the non-null fields of section~\ref{sub_pp_nonnull}). Namely, Ricci-flat Kundt metrics of Weyl type~N also solve Lagrangian theories ${ L}={ L}(g_{ab},R_{abcd},\nabla_{a_1}R_{bcde},\dots,\nabla_{a_1\dots a_p}R_{bcde})$ constructed from arbitrary powers of the Riemann tensor and its covariant derivatives and are thus ``universal'' \cite{HerPraPra14} (earlier results in special cases include \cite{Deser75,Guven87,AmaKli89,HorSte90,Horowitz90}). Similarly, their single-copied null electromagnetic fields are immune to corrections expressed in terms of arbitrary powers and derivatives of the field strength \cite{OrtPra16,OrtPra18,HerOrtPra18} (see also the earlier works \cite{Schroedinger35,Schroedinger43,Deser75,Guven87}).\footnote{Bearing in mind possible pathologies of null fields in particular theories such as ModMax electrodynamics \cite{Bandosetal20}.} However, note that higher-order theories may admit also additional solutions on top of the universal ones (e.g., non-Ricci-flat metrics or non-Maxwellian electromagnetic fields), for which a possible double copy would need to be studied on a case-by-case basis. 

{It is worth mentioning that, after the KS approach of \cite{MonOCoWhi14}, a different formulation of the classical double copy has been put forward in \cite{Lunaetal19} (see also the early results \cite{WalPen70,Hughstoneetal72}), namely the Weyl double copy. The latter has been established for all twistfree type N vacua in four dimensions in \cite{Godazgaretal21}, thus in particular for Kundt metrics of type N (see also \cite{Lunaetal19} for the special case of \pp waves). A natural question to ask is whether this result extends to higher dimensions. However, since the set-up of \cite{Lunaetal19} relies heavily on using 2-spinors in four-dimensional spacetime,  this issue will deserve further investigation (see \cite{MonNicOco19} for some results in this direction).}

This paper is organized as follows. 
Section~\ref{sec_prelim} contains some preliminaries. In section~\ref{sec_eqs}, we express the Einstein and Maxwell equations in a coordinate-free manner using the higher-dimensional NP formalism, derive the double-copy compatibility conditions, and discuss a special gauge simplifying these equations. Section~\ref{sec_KundtNP} establishes the double copy for Kundt waves. The main part of the proof is presented using the NP formalism, however, a result concerning the existence of a certain preferred null frame needed for the proof in the NP approach is obtained using the Kundt coordinates.   Section~\ref{sec_pp} is devoted to \pp waves. In particular, the existence of both null and non-null single copies is shown using the NP formalism. For completeness, a discussion in the Kundt coordinates is also included.  Finally, section~\ref{sec_nonunique} returns to the discussion of the non-uniqueness of the single copy encountered already, e.g., in section~\ref{sec_pp}.  

Various auxiliary results are presented in the appendices. Appendix~\ref{app_NP} contains a concise summary of the higher-dimensional Newman-Penrose formalism used throughout the paper.
Appendix~\ref{app_Riem} presents the Ricci rotation coefficients and the curvature for KS-Kundt spacetimes. Appendix~\ref{app_debney} extends  the useful coordinate system and frame defined for $n=4$ KS-Kundt spacetimes in \cite{Debney73,Debney74} to arbitrary dimension, and further discusses their role in connection to the general analysis of sections~\ref{sec_KundtNP} and \ref{sec_pp}. Finally, in Appendix \ref{app_example} we illustrate the above mentioned ``universal'' character of the KS-Kundt double copy  using specific examples of  modified theories of gravity and electromagnetism.

\section{Preliminaries }

\label{sec_prelim}

We define a frame adapted to the KS ansatz, i.e., a set of $n$ vectors $\bm_{(a)} $ which consists of two null vectors $\bl\equiv\bm_{(0)}$,  $\bn\equiv\bm_{(1)}$ and $n-2$ orthonormal spacelike vectors $\bm_{(i)} $, with $a, b\ldots=0,\ldots,n-1$, while $i, j  \ldots=2,\ldots,n-1$ \cite{Coleyetal04,OrtPraPra13rev}, such that $g_{ab}=\ell_a n_b+n_a \ell_b+m^{(i)}_am^{(i)}_b$.\footnote{With a slight abuse of notation, we will use the symbol $\bl$ for both the vector field $\ell^a\pa_a$ and the corresponding covector (1-form) $\ell_a\d x^a$ (where $\ell_a=g_{ab}\ell^b$). It will be clear from the context what is the object being considered.} The Ricci rotation coefficients $L_{ab}$, $N_{ab}$ and $\M{i}{a}{b}$ are defined by \cite{Pravdaetal04}
\be
 L_{ab}\equiv\ell_{a;b} , \qquad N_{ab}\equiv n_{a;b}  , \qquad \M{i}{a}{b}\equiv m^{(i)}_{a;b} ,
 \label{Ricci_rot}
\ee
and the corresponding frame components satisfy the identities 
\be 
L_{0a}=N_{1a}=N_{0a}+L_{1a}=\M{i}{0}{a} + L_{ia} = \M{i}{1}{a}+N_{ia}=\M{i}{j}{a}+\M{j}{i}{a}=0.
\ee

Since $\bl$ is geodesic (cf. above), without loss of generality we may take it to be affinely parametrized  (which is equivalent to setting  $L_{10}$ to zero by a boost, see \eqref{boosts} and \eqref{boost_coef}) with an affine parameter $r$ such that
	\be
	 \ell^a\pa_a=\pa_r .
	 \label{affine}
	\ee
Then employing a frame parallelly transported along $\bl$, one has 
\be
L_{i0}=L_{10}=\M{i}{j}{0}=N_{i0}=0.
 \label{pt}
\ee
This leaves the freedom of spins~\eqref{spins}, null rotation about $\bl$~\eqref{nullrot} and boosts~\eqref{boosts} (cf.~\cite{Coleyetal04}), provided all the transformation parameters are constant along $\bl$. 

The {\em optical matrix} is defined as 
	\be
	L_{ij}\equiv\ell_{a;b}m_{(i)}^am_{(j)}^b , 
	\label{L_def}
	\ee
see, e.g., \cite{Pravdaetal04,OrtPraPra13rev}. In particular, the null congruence defined by $\bl$ is expansionfree, twistfree, and shearfree precisely when $L_{ij}=0$, which characterizes the Kundt class of spacetimes.

Covariant derivatives along the frame vectors are denoted as
\be
	D \equiv \ell^a \nabla_a, \qquad \Delta\equiv n^a \nabla_a, \qquad \delta_i \equiv m_{(i)}^{a} \nabla_a,  
 \label{covder}
\ee
so that
\be
	\nabla_a=l_a \Delta +n_a D+m^{(i)}_a\delta_i, \label{der_f}
\ee 
and their commutators are given in \eqref{com_TD}--\eqref{com_dd}.

The box operator (acting on scalar functions) can be written as
\beqn
 \Box & & \equiv \nabla^a\nabla_a=\Delta D +  D\Delta + \delta_i\delta_i+L_{ii}\Delta+(N_{ii}-L_{11})D+(\MM{i}{jj}-L_{i1})\delta_i \nonumber \\
		  & & =2\Delta D  + \delta_i\delta_i+L_{ii}\Delta+(N_{ii}-2L_{11})D+(\MM{i}{jj}-2L_{i1})\delta_i , \label{box}
\eeqn	
where we used $L_{10}=0=N_{i0}$ (cf. \eqref{pt}) and, in the second equality, the commutator~\eqref{com_TD}.

For the KS metric \eqref{KS}, the null vector $\bl$ is not unique since it can be rescaled together with the metric function $\H$ without changing the metric ($\H \ell_a \ell_b =\hat \H \hat \ell_a\hat \ell_b$, where $\hat \ell_a= \lambda \ell_a$, $\hat \H=\lambda^{-2} \H$). The rescaling of $\bl$ corresponds to  a boost 
	\be
	\hbl = \lambda \bl, \qquad  \hbn = \lambda^{-1} \bn, \qquad \hbm{i} = \bmi{i} ,
	\label{boosts}
	\ee
	with $\lambda$ being a real function. Ricci rotation coefficients transform under the boost \eqref{boosts} according to~\eqref{boost_coef}, and we need to require $D\lambda=0$ if we want to preserve the condition $L_{10}=0$.

Throughout the paper, we will consider only KS spacetimes for which the KS vector field $\bl$ is Kundt, i.e., from now on we assume
	\be
	L_{ij}=0 ,
	\label{kundt}
	\ee
along with~\eqref{pt}. Under~\eqref{kundt}, the operator~\eqref{box} can thus be rewritten compactly as 
\be
 \Box f =\tilde\Box f +2D\big(\H Df) 
 \label{box2}
\ee	
for an arbitrary function $f$, where $\tilde\Box$ is the box operator associated to the flat metric $\bet$. (Eq.~\eqref{box2} follows readily from the comments in appendix~\ref{app_Riem}, cf.~\eqref{covder_KS}--\eqref{L11_KS}.) In particular, on a function $f$ which satisfies $Df=0$, one clearly has $\Box f=\tilde\Box f$, which will be useful for later purposes. Cf. also \cite{Debney73} in four dimensions.

\section{Field equations and double-copy compatibility conditions }

\label{sec_eqs}

In this section, we work out the Einstein and Maxwell equations for KS-Kundt solutions in the Newman-Penrose formalism, as well as their compatibility conditions in the set-up of the double copy (as described above in section~\ref{intro}). We also briefly discuss a non-uniqueness of the KS vector field related to  boost freedom (further comments will be given in section~\ref{sec_nonunique}), and use it to identify a particular gauge that will be convenient for later purposes. These results will be employed in the subsequent sections~\ref{sec_KundtNP} and \ref{sec_pp}.

\subsection{Field equations}

In the parallelly transported frame defined above obeying~\eqref{pt} and \eqref{kundt}, the components of the vacuum Einstein equations $R_{ab}=0$ read (cf.~\eqref{R01}--\eqref{R11})
\beqn
  & & \hspace{-.7cm} D^2 \H = 0 , \label{E01_K} \\
	& & \hspace{-.7cm}  (\delta_i D - 2 L_{[i1]}D)\H =0   ,  \label{E1i_K} \\
	& & \hspace{-.7cm} \Big[\delta_i\delta_i +  N_{ii}   D + ( 4 L_{1i} - 2 L_{i1} + \MM{i}{jj} ) \delta_i  \Big]\H	+2  \H \Big( 2 \delta_i L_{[1i]} + 4 L_{1i} L_{[1i]} + L_{i1} L_{i1}
               + 2 L_{[1i]} \MM{i}{jj}\Big)=0 . 
							\label{E11_K}
\eeqn

For the Maxwell field $\bF=\d\bA$ we take the ansatz 
\be
	\bA=\alpha\bl ,
	\label{A}
\ee	
where $\alpha$ is a spacetime function. The non-zero components of $\bF$ thus are 
\be
F_{01}=D\alpha, \qquad  
 F_{1i}=-2\alpha L_{[1i]} - \delta_{i}\alpha.\label{elmgF}
 \ee
  The field $\bF$ is {\em null} (cf., e.g., \cite{Sokolowskietal93,Coleyetal04vsi,OrtPra16}) iff $D\alpha=0 $.

The frame components of the Maxwell equations $(\sqrt{-g}F^{ab})_{,b}=0$ take the form (cf. also~\cite{Durkeeetal10,Ortaggio14})
\bea
  & & D^2\alpha=0	,		\label{M0_K} \\								
  & &  \left(\delta_i D-2L_{i1}D\right)\alpha  =0  ,
 		\label{Mi_K} \\		
  & &   \Big(\Delta D + \delta_j \delta_j+4 L_{[1j]}\delta_{j}+ N_{jj}D + \overset{k}{M}_{jj}\delta_{k}\Big) \alpha  +2\alpha\left(2L_{[1j]}L_{[1j]}+\delta_{j}L_{[1j]}+ L_{[1k]}\overset{k}{M}_{jj} \right) =0 ,
				\label{M1_K} 		
\eea
where in~\eqref{Mi_K} we have used the Ricci identity \eqref{Ricci_dL} and the commutator \eqref{com_Tdi}.

Let us start with the Einstein equations. Eq.~\eqref{E01_K} gives
\be
	\H=r\H^{(1)}+\H^{(0)} ,
	 \label{H_K}
\ee	
where $r$ is defined in~\eqref{affine}, and $D\H^{(1)}=0=D\H^{(0)}$. For later purposes, it is also useful to observe that (using the commutator~\eqref{com_Tdi} and a suitable null rotation~\eqref{nullrot} leaving $\bl$ unchanged, cf. \cite{PraPra08}),  one can set 
\be
 \delta_ir=-L_{1i}r .
 \label{delta_r}
\ee

The remaining Einstein equations~\eqref{E1i_K}, \eqref{E11_K} become\footnote{To be precise, \eqref{E11_K} contains a term independent of $r$, which gives rise to \eqref{E11_K2}, and a term linear in $r$, which can be shown to be identically zero (cf., e.g., \cite{OrtPra16}). The fact that all the quantities entering~\eqref{E11_K2} are $r$-independent follows from \eqref{Ricci_dL}, \eqref{Ricci_DN}.}
\beqn
  & & \delta_i \H^{(1)}=2L_{[i1]}\H^{(1)} , \label{E1i_K2} \\
	& & \Big[\Box+2L_{1i}\Big(2\delta_i +4L_{[1i]}+\MM{i}{jj}\Big)+2(\delta_iL_{1i})\Big]\H^{(0)}+N_{ii}\H^{(1)}=0 , 
							\label{E11_K2}
\eeqn
where in~\eqref{E11_K2} we have used~\eqref{box}, \eqref{E1i_K2} and \eqref{Ricci_dL}, \eqref{Ricci_DN}. Recall that (cf.~\eqref{box2})
\be
 \Box\H^{(0)}=\tilde\Box\H^{(0)} , 
\ee	
i.e., the wave operator in curved spacetime reduces to its Minkowskian counterpart when acting on $\H^{(0)}$.

The Maxwell equations will be discussed in the next subsection in the form of a set of ``compatibility conditions'', motivated by the KS double-copy construction \cite{MonOCoWhi14}.

\subsection{Double-copy compatibility conditions}

\label{subsec_double}

Following the double-copy prescription \cite{MonOCoWhi14} discussed above  (eqs.~\eqref{KS} and \eqref{A_dc}), from now on in~\eqref{A} we set
\be
 \alpha=\H .
 \label{a=H}
\ee
We aim to determine a set of compatibility conditions such that the Maxwell equations~\eqref{M0_K}--\eqref{M1_K} with~\eqref{a=H} become a consequence of the Einstein equations~\eqref{E01_K}--\eqref{E11_K} (i.e., \eqref{H_K}, \eqref{E1i_K2}, \eqref{E11_K2}).

While equations~\eqref{E01_K} and \eqref{M0_K} are identical, by comparing~\eqref{E1i_K} with \eqref{Mi_K} and~\eqref{E11_K} with \eqref{M1_K}, we obtain the following compatibility conditions
\beqn
 & & (L_{1i}+L_{i1})D\H=0 , \label{compi_K} \\
 & & \Big[\Delta D  -L_{1i}(2\delta_i+3L_{1i}-2L_{i1}+\overset{i}{M}_{jj})-(\delta_iL_{1i})
 \Big]\H=0 ,
				\label{comp1_K} 		
\eeqn
where in~\eqref{comp1_K} we have used the Ricci identity \eqref{Ricci_dL} to get rid of a term proportional to $\delta_iL_{i1}$. 

Using~\eqref{H_K} and \eqref{E1i_K2}, we can rewrite the compatibility conditions~\eqref{compi_K} and \eqref{comp1_K} as
\beqn
 & & (L_{1i}+L_{i1})\H^{(1)}=0 , \label{compi_K2} \\
 & & \H^{(1)} 
 (L_{1i}-\overset{i}{M}_{jj}-\delta_i)L_{1i}=0 , \label{comp1_K2a} \\ 
 & & \Delta\H^{(1)}-\left[L_{1i}(3L_{1i}-2L_{i1}+\overset{i}{M}_{jj}+2\delta_i)+(\delta_i L_{1i})
  \right]\H^{(0)}=0 . \label{comp1_K2b} 
\eeqn

Let us observe that, if $\H^{(1)}\neq0$, the compatibility conditions~\eqref{compi_K2} and \eqref{comp1_K2a} with the Ricci identity \eqref{Ricci_dL} lead to $L_{1i}=0=L_{i1}$. This case corresponds to a subset of \pp waves discussed in section~\ref{sub_pp_nonnull}, while in all the remaining cases $\H^{(1)}=0$.

 Recalling that the quantities $L_{i1}$ are invariant under boosts (while the $L_{1i}$ are not, cf.~\eqref{boosts2}), it will be useful to study separately the two invariantly defined cases of ``Kundt waves'', defined by $L_{i1}\neq0$, and of ``\pp waves'', corresponding to $L_{i1}=0$ (cf. also \cite{OrtPraPra13rev} and references therein) in sections~\ref{sec_KundtNP} and \ref{sec_pp}, respectively.

\subsection{Boost freedom and gauge $L_{1i}=0$}

\label{subsec_gauge}

\subsubsection{Boost freedom}

\label{subsubsec_boost}

Before proceeding, it is useful to comment on the gauge freedom mentioned in sections~\ref{intro} and \ref{sec_prelim}. Namely, one can  perform a boost~\eqref{boosts} of the vector $\bl$ with $D\lambda=0$  and a simultaneous rescaling of $\H$ in such a way that the KS metric remains unchanged, along with the conditions~\eqref{pt}, see \eqref{boost_coef}. Naturally, the Einstein equations~\eqref{E01_K}--\eqref{E11_K} are invariant under this boost-rescaling transformation. In contrast, the Maxwell equations~\eqref{Mi_K}, \eqref{M1_K} are not invariant under this transformation (whilst~\eqref{M0_K} is). The electromagnetic potential changes as $A_a=\H \ell_a\ \rightarrow\ \hat A_a=\hat \H\hat \ell_a=\lambda^{-1} A_a$. Demanding that both $\bA$ and $\hat\bA$ obey the Maxwell equations leads to a differential condition containing $\lambda$ and its (up to 2nd) derivatives  (cf. section~\ref{nonuni1}).  Therefore, in principle,  distinct electromagnetic fields could be related to a given KS metric via the KS double copy, see the discussion in section~\ref{sec_nonunique}. See also \cite{CarPenTor18} for related comments.

\subsubsection{Gauge $L_{1i}=0$}

\label{subsubsec_gauge}

Here we will use the boost freedom described above to identify a particularly convenient gauge. Namely, given an affinely parametrized Kundt vector field $\bl$, one can always use a boost~\eqref{boosts} with $D\lambda=0$ to set $\hat L_{1i}=0$, and thus also $\delta_i\hat r=0$, cf.~\eqref{delta_r} (the Ricci identity for $\delta_{[j|}L_{1|i]}$ \eqref{Ricci_dL}  and the commutator \eqref{com_dd} ensure  that the necessary integrability conditions are satisfied; one should also redefine $\hat r=\lambda^{-1}r$  in order to have $\hat \bl=\pa_{\hat r}$). This can be done at once both for the background spacetime (i.e., for $\H=0$) and for the full metric \cite{OrtPraPra09} (cf.~\eqref{Nij_KS}). There still remains residual boost freedom with $D\lambda=0=\delta_i\lambda$. In the rest of this section, this boosted frame will be understood and hats above boosted quantities will thus be dropped.

With the gauge choice $L_{1i}=0$, the compatibility conditions~\eqref{compi_K2}--\eqref{comp1_K2b} reduce to 
\beqn
 & & L_{i1}\H^{(1)}=0 , \label{compi_K3} \\
 & & \Delta\H^{(1)}=0 , \label{comp1_K3} 
\eeqn
while the Einstein equations~\eqref{E1i_K2} and \eqref{E11_K2} become
\beqn
  & & \delta_i \H^{(1)}=L_{i1}\H^{(1)} , \label{E1i_K3} \\
	& & \Box\H^{(0)}+N_{ii}\H^{(1)}=0 . \label{E11_K3}
\eeqn

This gauge will be chosen in sections~\ref{sec_KundtNP} and \ref{subsec_pp_gauge}  below.

\section{Case $L_{i1}\neq0$: higher-dimensional Kundt waves}

In this section, we prove the KS double copy for Kundt waves in arbitrary dimension. First, we present a general, coordinate-independent analysis in the Newman-Penrose formalism (section~\ref{subsec_KundtNP_gen}), building on the set-up of section~\ref{sec_eqs}. We next give a more explicit demonstration of how the double copy is realized in the canonical Kundt coordinates of \cite{Kundt61,Kundt62,Coleyetal06} (section~\ref{Kundt_coords}).

\label{sec_KundtNP}

\subsection{Proof of the KS double copy for all Ricci-flat Kundt waves }

\label{subsec_KundtNP_gen}

It follows from the compatibility conditions (see \eqref{compi_K2}) and the comments given at the end of section~\ref{subsec_double}, that here
\be
 \H^{(1)}=0 . 
 \label{H1_Kundt}
\ee

The compatibility conditions \eqref{compi_K2}--\eqref{comp1_K2b} thus reduce to 
\be
   \left[L_{1i}(3L_{1i}-2L_{i1}+\overset{i}{M}_{jj}+2\delta_i)+(\delta_iL_{1i})
   \right]\H^{(0)}=0 , \label{comp1_K_Kundtw} 
\ee
while the Einstein equations \eqref{E1i_K2}, \eqref{E11_K2} become simply 
\be
  \Big[\Box+2L_{1i}\Big(2\delta_i +4L_{[1i]}+\MM{i}{jj}\Big)+2(\delta_iL_{1i})\Big]\H^{(0)}=0 . 
							\label{E11_K_Kundtw}
\ee
Using~\eqref{comp1_K_Kundtw}, the latter can be rewritten as $\left(\Box-2L_{1i}L_{1i} 
\right)\H^{(0)}=0$.

From now on we choose the gauge (cf. section~\ref{subsec_gauge})
\be
 L_{1i}=0 .\label{L1i0}
\ee 
(This can be done without affecting~\eqref{H1_Kundt}.)
Hence  the compatibility eq.~\eqref{comp1_K_Kundtw} becomes an identity and the Einstein eq.~\eqref{E11_K_Kundtw} reduces to
\be
\Box \H^{(0)}=0 . \label{Kundt_EFE}
\ee

Since the compatibility conditions are now satisfied identically, this demonstrates the double copy for Kundt waves with $\H^{(1)}=0$ (eq.~\eqref{H1_Kundt}). Recall that this condition followed from the Einstein equations together with the compatibility conditions and the Kundt waves property $L_{i1}\not=0$. 

It thus remains to clarify whether the KS double copy holds for general type~N vacuum Kundt waves (that could have, in principle, $\H^{(1)}\neq0$). Indeed, one can argue that $\H^{(1)}$ does not contribute to the spacetime curvature and can thus always be reabsorbed into the $\bet$ part of the metric or transformed away by a coordinate transformation.  This is illustrated in section~\ref{subsubsec_pt2} (cf. eq.~\eqref{phi_frame2}) and appendix~\ref{app_debney} (eqs.~\eqref{KS_debney2}, \eqref{KS_debney2_back}). Therefore, we conclude that the KS double copy holds for all Ricci flat type N Kundt waves. An independent, coordinate-based proof will be given in section~\ref{sec_Kundtline}.

Let us further note that the residual boost freedom with $D\lambda=0=\delta_i\lambda$ mentioned in section~\ref{subsubsec_gauge} gives rise to a special instance of the case~1. of the non-uniqueness discussed in~section~\ref{nonuni1}.

\subsection{Kundt waves in Kundt coordinates}

\label{Kundt_coords}

This section focuses on analyzing the KS double copy for Kundt waves using the canonical Kundt coordinates. For some applications, this may be a useful addition
to the discussion in Newman-Penrose formalism given above.

\subsubsection{Line-element and double copy}

\label{sec_Kundtline}

All Weyl type N Ricci-flat Kundt metrics belong to the so-called VSI (vanishing scalar invariants) class of spacetimes \cite{Coleyetal04vsi} admitting a metric of the form \cite{Coleyetal06}  
\be
\d s^2 =2\d u\left[\d v+H(u,v,x^k)\d u+W_{ i}(u,v,x^k)\d x^{ i}\right]+ \delta_{ij} \d x^i\d x^j , 
\label{VSI}
\ee
where $i,j=2,\ldots,n-1$ and
\bea
W_{i}(u,v,x^k)&=&-\delta_{i, 2} \frac{2 \epsilon}{\y} v + W^{(0)}_{i} (u,x^k), \label{VSIW}\\
H(u,v,x^k)& = &\frac{\epsilon v^2}{2 \y^2}+ v H^{(1)} (u,x^k) + H^{(0)} (u,x^k), \qquad \epsilon=0,\ 1,\qquad \y\equiv x^2. \label{VSIH}
\eea
The canonical Kundt covector field is given by $\ell_a\d x^a=\d u$, for which $\ell_{a;b}=\left(\frac{\epsilon v}{y^2}+H^{(1)}\right)\d u^2-\frac{\epsilon}{y}(\d u\d y+\d y\d u)$  \cite{Coleyetal06,OrtPra16}.

Consider a null electromagnetic field $\bF=\d\bA$ with 
	\be 
	\bA=\alpha(u,x^k)\d u, \qquad 
	\bF=f_i\d u\wedge \d x^i,\quad 
	f_i= F_{ui}=-\alpha,_i.\label{potA} 
\ee 
Then the Maxwell equations  in any spacetime~\eqref{VSI}  reduce to \cite{OrtPra16}
\be 
\left( \sqrt{-\mbox{det}g_{ab}} f^i\right),_i
=\left( \sqrt{\mbox{det} g_{ij} }f^i\right),_i=f^i,_i=-\delta^{ij}\alpha,_{ij}=
-\triangle\alpha=0,\label{Maxwell} 
\ee 
where $\triangle$ stands for the Euclidean Laplace operator in the $(n-2)$-dimensional flat space spanned by the coordinates $x^i$ (not to be confused with the Newman--Penrose symbol $\Delta$ defined in~\eqref{covder}).\footnote{Note that, for any $v$-independent scalar function $\alpha$, one has $\Box\alpha=\triangle\alpha+\frac{2\epsilon}{y}\alpha_{,y}$.}

The VSI metrics \eqref{VSI}--\eqref{VSIH} are in general of Weyl and Ricci type III and thus contain also some non-KS spacetimes. However, here we are interested in Kundt waves, i.e., in the Weyl type N, Ricci-flat subclass of metrics~\eqref{VSI}--\eqref{VSIH} with $L_{i1}\neq0$. This is given by the family $\epsilon=1$ for which, after using some of the Einstein equations and  coordinate freedom, eqs.~\eqref{VSIW} and \eqref{VSIH} simplify to \cite{Coleyetal06} (see also \cite{Coleyetal03})
\begin{eqnarray}
	W_{2} & = & -2\frac{v}{\y},  \label{VSI_K_W2}\\
	W_{m} & = & x^{n}B_{nm}(u)+C_{m}(u) \qquad (m,n,p,q=3,\dots,n-1), \label{VSI_K_Wm}\\
	H & = & \frac{v^2}{2\y^{2}}+H^{(0)}(u,x^{i}),\label{VSI_K_H}  
\end{eqnarray}
where $C_m$ and $B_{nm}=-B_{mn}$ are arbitrary functions of $u$ (in the special case $n=4$, one thus has $B_{nm}=0$, corresponding to the canonical form of \cite{Kundt61}).

The only non-trivial component of the Einstein vacuum equations reads
	\begin{equation}
		\y \bigtriangleup \left( \frac{H^{(0)}}{\y} \right)-\frac{1}{\y^2}  W_m W_m - B_{mn} B_{mn}=0. \label{eqEinstein_Kundt}
\end{equation} 
The metric~\eqref{VSI} with \eqref{VSI_K_W2}--\eqref{VSI_K_H} is flat for $H$ given in \eqref{VSI_K_H} with
\be
		H^{(0)}(u,x^{i})=H^{(0)}_{\mbox{\tiny flat}} =
\frac{1}{2} W_mW_m +\y F_0(u)+\y x^iF_{i}(u).
 \label{H0_Kundt}
\ee

Now defining $\H^{(0)} $ 
\be
\H^{(0)} =\ff \y\equiv  H^{(0)}_{\mbox{\tiny flat}}-H^{(0)},   
\ee
the metric takes the KS form
\be
\d s^2 =\d s^2_{\mbox{\tiny flat}} -2  \H^{(0)} \d u^2 ,\label{eqKundt_KS}
\ee
where $D \H^{(0)} =0$.
Since $H^{(0)}_{\mbox{\tiny flat}}$ solves \eqref{eqEinstein_Kundt}, the Einstein equation \eqref{eqEinstein_Kundt} reduces to
\be
\triangle \ff =0.\label{Einstein_Kundt}
\ee
Now rewriting the metric \eqref{eqKundt_KS} in the form 
\be
\d s^2 =\d s^2_{\mbox{\tiny flat}}  -2 \y\ff \d u^2
=\d s^2_{\mbox{\tiny flat}} {-2} \hat\H  \hat\bl\otimes\hat\bl ,
 \label{kundtw_db}
\ee
where $\hat\H=\ff /\y$ and $\hat\bl  = \y \d u$ and
taking potential $\bA=\hat\H \hat\ell={\ff} \d u$ \eqref{potA} ($A_u={\ff}=A^v$), 
the Maxwell equations \eqref{Maxwell} both in the curved \eqref{VSI_K_W2}--\eqref{VSI_K_H} and flat backgrounds, 
\be
\triangle  \ff =0,
\ee
are equivalent to the Einstein equations~\eqref{Einstein_Kundt}, which establishes the KS double copy for higher-dimensional Kundt waves. An equivalent observation in four dimensions was made some time ago \cite{Kundt61,Siklos85}.

Since the standard natural null frame for Kundt metrics \eqref{VSI} \cite{Coleyetal06,OrtPraPra13rev},
\bea
\ell_a\d x^a&=&\d u\quad  (\Leftrightarrow\ \ell^a\pa_a=\partial_v),\label{naturframe_l}\\
n_a\d x^a&=&\d v+H\d u+W_i\d x^i,\label{naturframe_n}\\
m^{(i)}_a\d x^a&=&\d x^i,\label{naturframe_mi}
\eea
  is not, for Kundt waves, parallelly transported along $\bl$ (since it gives $N_{i0}\neq0$), let us conclude this section by presenting two frames that are parallelly propagated along $\bl$. This will enable us to make contact with the analysis of section~\ref{subsec_KundtNP_gen}.

To conclude this section, let us observe that, without loss of generality, one can in fact set $W_{m}=0$ in~\eqref{VSI} and \eqref{VSI_K_Wm}: $C_m$ can be set to zero by redefining $x^m\mapsto x^m+f^m(u)$ with $f^m,_u=-C_m$, while $B_{nm}$ can be set to zero with a rotation  $x^m\mapsto {\R^m}_n (u)x^n$, $\delta_{pq}{\R^p}_m {\R^q}_n=\delta_{mn}$ with $ {\R^p}_{m,u}=B_{pn}{\R^n}_m$. This observation seems to have been overlooked in the literature so far.  Under such a coordinate fixing, formulae~\eqref{eqEinstein_Kundt}, \eqref{H0_Kundt}, and \eqref{naturframe_n} simplify accordingly.

\subsubsection{{Parallelly transported frame  with $L_{10}=0\neq L_{1i}=L_{i1}$}}	

By a null rotation~\eqref{nullrot} (with $z_2=v/y$, $z_m=0$) of the frame \eqref{naturframe_l}--\eqref{naturframe_mi}, one obtains a parallelly transported frame (i.e., $L_{i0}=L_{10}=N_{i0}=\MM{i}{j0}=0$, cf.~\eqref{pt})\footnote{To keep the notation simple, hats on transformed quantities will be omitted from now on.} \bea
& & \bl_a \d x^a = \d u, \qquad 
n_a\d x^a=  \d v+H_0 \d u  - \frac{v}{\y} \d \y+W_m\d x^m,
\label{frame_11}\\
& & m^{(2)}_a \d x^a = -\frac{v}{\y} \d u +  \d \y, \qquad m^{(m)}_a \d x^a =   \d x^{m} , \label{frame_12}
\eea
for which 
\be
	 L_{11}=-\frac{v}{y^2} , \qquad  \MM{i}{jk}=0,   \qquad N_{ii}=0, \qquad  L_{12}=L_{21}=-\frac{1}{y},
	\qquad  L_{1m}=L_{m1}=0.
\ee

Since $L_{12}\not=0$ this frame clearly does not correspond to the gauge $L_{1i}=0$ employed in section~\ref{subsubsec_gauge}, which we implement in the next step.

\subsubsection{{Parallelly transported frame  with $L_{10}=L_{1i}=0\neq L_{i1}$}}	

\label{subsubsec_pt2}

By boosting (eq.~\eqref{boosts} with $\lambda=y$) the above frame~\eqref{frame_11}, \eqref{frame_12},  one obtains a parallelly transported frame
\bea
 & &  \bl_a \d x^a =\y \d u, \qquad
 n_a\d x^a= \frac{1}{\y} \left( \d v+H_0 \d u - \frac{v}{\y} \d \y+W_m\d x^m
\right) ,\label{frame_21}\\
& &  m^{(2)}_a \d x^a = -\frac{v}{\y} \d u +  \d \y, \qquad
m^{(m)}_a \d x^a =   \d x^{m}	, \label{frame_22}
\eea
for which
\be
L_{10}=0,  \qquad  L_{11}=0, \qquad  \MM{i}{jk}=0,    \qquad N_{ii}=0, \qquad  L_{21}=-\frac{1}{y},
\qquad  L_{1m}=L_{m1}=L_{12}=0.
\ee

The existence of the frame \eqref{frame_21}--\eqref{frame_22} has been employed in section~\ref{subsec_KundtNP_gen} (for four-dimensional Kundt waves, a parallelly transported frame with $L_{1i}=0$ was also constructed in \cite{Debney74} using different coordinates, see appendix~\ref{app_debney}).

A further null rotation with $z_2=0$, $z_m=-W_m/y$ accompanied by the coordinate transformation 
\be
 u'=u , \qquad r=\frac{v}{y} , \qquad y'=y , \qquad x'^m=x^m , \qquad 
 \label{r_v}
\ee
brings the derivative operators to the form (dropping hats on the operators in the new frame)
\be
 D=\pa_r , \quad \Delta=\frac{1}{y'}\left[\pa_{u'}+\frac{1}{y'}\left(-H^{(0)}+\frac{1}{2}W_mW_m\right)\pa_r+r\pa_{y'}-W_m\pa_{x'^m}\right] , \quad  \delta_2=\pa_{y'} , \quad \delta_m=\pa_{x'^m}  .
\ee
which means that 
\be
 \H=\frac{1}{(y')^2}(H^{(0)} - H^{(0)}_{\mbox{\tiny flat}}) ,
 \label{phi_frame2}
\ee 
in agreement with~\eqref{kundtw_db}. This gives an explicit illustration of the parallelly transported frame introduced in a coordinate-independent way in section~\ref{subsec_KundtNP_gen} and proves that for all Kundt waves, one can set $\H^{(1)}=0$ while using such a frame.

\section{Case $L_{i1}=0$: higher-dimensional  \pp waves}

\label{sec_pp}

Here $\ell_{a ; b }=\ell_a(L_{11}\ell_b +L_{1i}\msub{i}{b})$ (cf.~\eqref{dl_kundt}), i.e., $\bl$ is a recurrent vector field \cite{Stephanibook,OrtPraPra13rev}. It is not difficult to see that $\bl$ is, in fact, always proportional to a covariantly constant null vector field\footnote{First, an appropriate boost $\hbl = \lambda \bl$~\eqref{boosts} can always be made which sets $\hat L_{1i}=0$ in the new frame (cf. \cite{Kundt61,DebCah61,KerGol61,GibPop08} and section \ref{subsubsec_gauge}); next, the Ricci identities \eqref{Ricci_DN} and \eqref{Ricci_deltaL11} give  $D\hat L_{11}=0=\delta_i\hat L_{11}$, which means one can perform a further boost $\tbl= \eta\hbl$ with $D\eta=0=\delta_i\eta$, $\frac{\bigtriangleup \eta}{\eta}=\hat L_{11}$, and thus set $\tilde L_{11}=0$ in the final frame (while keeping~\eqref{pt} and $\tilde L_{1i}=0=\tilde L_{i1}$) -- i.e., $\tilde \ell_{a ; b }=0$.} and therefore the corresponding spacetime is a \pp wave. In contrast to the case of Kundt waves of section~\ref{sec_KundtNP}, now two cases need to be studied separately, depending on whether $\H^{(1)}$ vanishes or not.

\subsection{Case $\H^{(1)}=0$: null electromagnetic field}

\label{subsec_pp_gauge}

Here the compatibility conditions~\eqref{compi_K2}--\eqref{comp1_K2b} become a single equation
\be
\left[L_{1i}(3L_{1i}+\overset{i}{M}_{jj}+2\delta_i)+(\delta_iL_{1i})\right]\H^{(0)}=0 , 
\label{comp1_K_pp}
\ee
and the Einstein equations~\eqref{E1i_K2}, \eqref{E11_K2} reduce to
\be
\Big[\Box+2L_{1i}\Big(2\delta_i +2L_{1i}+\MM{i}{jj}\Big)+2(\delta_iL_{1i})\Big]\H^{(0)}=0 .
\label{E11_K_pp}
\ee

\subsubsection{Gauge $L_{1i}=0$ }

\label{subsubsec_gauge_pp}

Using the gauge $L_{1i}=0$ (see section~\ref{subsec_gauge}), eq.~\eqref{comp1_K_pp} is satisfied identically and 
eq.~\eqref{E11_K_pp} gives simply
\be
\Box \H^{(0)}=0 , 
\ee
which demonstrates the {\em null} field double copy for all \pp waves (see also section~\ref{subsec_pp} for a coordinate approach and appendix~\ref{app_debney}; in particular, the arguments given at the end of section~\ref{subsec_KundtNP_gen} about the generality of the condition $\H^{(1)}=0$ apply also here).

\subsection{Case $\H^{(1)}\neq0$: non-null electromagnetic field }

\label{sub_pp_nonnull}

In this section, we consider the case $\H^{(1)}\neq0$ in \eqref{H_K}. Using \eqref{der_f},
we see that for the electromagnetic tensor $F_{ab}=A_b,_a-A_b,_a$, where $A_a=\H\ell_a $, the boost-weight zero component  $F_{01}=D\H=\H^{(1)}\not=0$ and thus $F_{ab}$ is non-null, cf. \eqref{elmgF}, \eqref{H_K}, and \eqref{a=H}.
The compatibility conditions \eqref{compi_K2}--\eqref{comp1_K2b} reduce to
\be
	L_{1i}=0, \qquad \Delta \H^{(1)}=0 , 
	\label{comp_pp}
\ee
while the Einstein equations~\eqref{E1i_K2}, \eqref{E11_K2} become
\beqn
  & & \delta_i \H^{(1)}=0 , \label{delta_phi_pp} \\
	& & \Box \H^{(0)}+N_{ii}\H^{(1)}=0 . \label{pp_H0}
\eeqn

Provided \eqref{comp_pp} is fulfilled, then the Einstein and Maxwell equations are equivalent. Since we already have $D\H^{(1)}=0$, it follows from the second of~\eqref{comp_pp} and \eqref{delta_phi_pp} that $\H^{(1)}$ must be a (non-zero) constant. It is proven in sections~\ref{subsec_pp}, \ref{sec_nonuni2}, and appendix~\ref{app_debney} that, for all vacuum \pp waves of type~N, one can indeed find a frame such that~\eqref{comp_pp} and $\H^{(1)}=\mbox{const}\not=0$ are both satisfied (while maintaining our earlier ``gauge fixings''~\eqref{pt} and \eqref{delta_r}; note that here the first of~\eqref{comp_pp} is not a gauge choice but a consequence of the compatibility condition \eqref{compi_K2}). This proves the {\em non-null}-field  KS double copy for \pp waves (to our knowledge, previous literature has considered only the case of null fields). An explicit example of a \pp wave metric and the corresponding non-null electromagnetic field in arbitrary dimension, using the Kundt coordinates, is given in section~\ref{sec_nonuni2}.

\subsection{\pp waves in Kundt coordinates}

\label{subsec_pp}

The canonical line-element in Kundt coordinates for Ricci-flat \pp waves of Weyl type N is given by~\eqref{VSI} with $\epsilon=0$ and \cite{Brinkmann25,Schimming74,Coleyetal06,Coleyetal03}\footnote{We note that the interesting simplification~\eqref{pp_W} was obtained in arbitrary dimensions in \cite{Schimming74} (see also \cite{Horowitz90,Sokolowski91},  footnote~5 of \cite{KucOrt19} and appendix \ref{app_debney}). For the special case $n=4$, this is well known \cite{Brinkmann25,EK,Kundt61,Stephanibook,GriPodbook}.}
\begin{eqnarray}
	W_i & = & 0 \label{pp_W} \\
	H & = & H^{(0)}(u,x^{i}) ,  \label{pp_H}   
\end{eqnarray}
which gives
\be 
\d s^2= 
2\d u \left[
\d v +H^{(0)}(u,x^k)\d u\right]
+ \delta_{ij} \d x^i\d x^j .
\label{VSIpp}
\ee

The remaining Einstein vacuum equation reads
\begin{equation}
\triangle H^{(0)}=0 .
\label{eqEinstein_PP}
\end{equation}

This metric is flat for  \cite{Coleyetal06}
\be
H^{(0)}(u,x^{i})=H^{(0)}_{\mbox{\tiny flat}}= G(u)+x^i F_i(u) ,  
\ee
where $G$ and $F_i$ are arbitrary functions of $u$ (they can be set to zero by a transformation $x^i\mapsto x^i+h_i(u)$, $v\mapsto v-\dot h_ix^i+g(u)$ \cite{Stephanibook,Coleyetal06}, if desired).

Now defining 
\be
\H= \H^{(0)} \equiv  H^{(0)}_{\mbox{\tiny flat}}-H^{(0)}
\label{HPP_KS}
\ee
the metric takes the KS form
\be
\d s^2 =\d s^2_{\mbox{\tiny flat}} -2\H^{(0)} \d u^2 . 
\label{eqPP_KS}
\ee
The  null vector $\bl=\d u$ is covariantly constant, $\nabla_a \bl_b  = 0$. (Here $\ell^a\pa_a=\pa_v$ and $L_{1i}=0=L_{i1}$ in any frame adapted to $\bl$, so in this case the coordinate $v$ can be identified with the affine parameter $r$~\eqref{affine} employed in sections~\ref{sec_eqs}, \ref{subsec_pp_gauge} and \ref{sub_pp_nonnull}).

$H^{0}_{\mbox{\tiny flat}}$ is a solution of \eqref{eqEinstein_PP} and thus the Einstein equation \eqref{eqEinstein_PP} reduces to
\be
\triangle \H^{(0)} =0. \label{Einsteq_pp}
\ee
Thus identifying $\H^{(0)}$ with $\alpha$ in \eqref{potA}, the Maxwell equations (both in flat and curved backgrounds) \eqref{Maxwell} are equivalent to the Einstein equations \eqref{Einsteq_pp}. This {establishes} the null-field KS double copy in the coordinate form. The same conclusion also follows from the test-field limit of the results of \cite{GurTek18} (cf. again \cite{Kundt61,Siklos85} for related results in four dimensions).

The null field KS double copy in the context of higher-dimensional $pp$-waves has been already discussed in \cite{MonOCoWhi14} in the coordinates~\eqref{VSIpp} with a fixed gauge $\bl=\d u$, in which $\bl$ is covariantly constant.

One can employ the coordinate transformation \eqref{hat_u} to generate a non-vanishing $H^{(1)}=H^{(1)}(u)$, see \eqref{hat_H}.
Since $H^{(1)}$ does not enter the field equations one can either identify it as a part of the flat background $H_{\mbox{\tiny flat}}$, see section \ref{subsec_pp_gauge}, or as a part of the KS function as in section \ref{sub_pp_nonnull}. The latter case is related to the non-null field KS double copy, see also section~\ref{sec_nonuni2}.

\section{Non-uniqueness of the KS double copy}

\label{sec_nonunique}

In section~\ref{sec_pp}, we have found that one gravitational field can have multiple distinct electromagnetic single copies. This was limited to \pp waves. Now let us discuss the non-uniqueness of electromagnetic single copies in more detail and generality.

	There are two distinct sources of non-uniqueness:
	\begin{enumerate}
	
		\item The first source of the non-uniqueness is tied to the non-uniqueness of the KS vector field $\bl$ under boosts~\eqref{boosts} discussed in sections \ref{sec_prelim} and \ref{sec_eqs}.
		While under the  boost and rescaling transformation $\H \ell_a \ell_b =\hat \H \hat \ell_a\hat \ell_b$, where 
		\be
		 \hat \ell_a= \lambda \ell_a , \qquad \hat \H=\lambda^{-2} \H , 
		 \label{rescal1}
		\ee
		the KS metric \eqref{KS} remains unchanged, the corresponding potential transforms as 
		\be
		 \hat\bA=\hat\phi\hat\bl={\lambda}^{-1} \bA .
		 \label{case1_A}
		\ee
		Distinct gauge fields can thus, in principle, correspond to the same metric via different choices of $\bl$ (but all defining the same null direction). If $\bF$ is null then $\hat\bF$ is null too, as long as $D\lambda=0$. 
		
		We shall show below (section~\ref{nonuni1}) that this kind of non-uniqueness applies to all Kundt and \pp waves, giving rise, in both cases, to a continuous infinity of (null-field) single copies.
		
		\item 
		The second source of the non-uniqueness is tied to the non-uniqueness of the splitting of the KS metric in the flat part and KS part of the metric~\eqref{KS}. 
		In some cases,  the KS metric~\eqref{KS} is flat not only for $\H=0$ but also for some non-trivial $\H=\tilde \H $ and then the term $\tilde \H \bl\otimes\bl$ can be either absorbed in the flat part of the metric $\d s^2_{\mbox{\tiny flat}}$ or left in the KS part $\H \bl\otimes\bl$ (see, e.g.,  \cite{Debney74,Urbantke79}  for related comments in four dimensions). In general, this can lead to distinct electromagnetic fields $\bA=\H \bl$, even if the same $\bl$ is used in all cases. In particular, the resulting field strength can be both null and non-null, depending on the chosen splitting. Below in section~\ref{sec_nonuni2}, we will illustrate this type of non-uniqueness in the context of \pp waves and establish its connection to the non-null-field single copy discussed in section~\ref{sub_pp_nonnull}.

	\end{enumerate}

\subsection{Case 1. -- infinite non-uniqueness of the single copy}

	\label{nonuni1}

In sections~\ref{sec_prelim}, \ref{subsubsec_boost}, and \ref{subsec_KundtNP_gen}, we pointed out a boost freedom with $D\lambda=0$ in the definition of the KS vector field $\bl$ and its corresponding metric function~$\H$ in~\eqref{KS}, which implies a non-uniqueness of the single-copy field in the double-copy procedure. Let us now show in detail how this non-uniqueness arises for both \pp waves and Kundt waves.

Let us begin with the type N Ricci flat \pp waves~\eqref{eqPP_KS} with the KS function $\H$ given	by~\eqref{HPP_KS}. Here $D\lambda=0$ means $\lambda=\lambda(u,x^i)$. When $\bF=\d \bA$, $\bA=\H\d u$ (a null-field single copy w.r.t. $\bl=\d u$, cf. also~\eqref{potA})\footnote{It is easy to see that in the case of the non-null-field single copy of \pp waves (section~\ref{sub_pp_nonnull}), i.e., of a KS solution \eqref{KS} with $\H=r\H^{(1)}+\H^{(0)}$ (eq.~\eqref{H_K}) and $\H^{(1)}\neq0$, the compatibility conditions~\eqref{compi_K2}--\eqref{comp1_K2b} for $\hat\phi=\lambda^{-2}\H$ imply (using~\eqref{boost_coef}) that $\lambda$ must be a constant, giving rise to a trivial rescaling of the electromagnetic field. For this reason we will not consider non-null electromagnetic fields in the present discussion.} satisfies the Maxwell field equations
\be
\phi,_{ii}=0 \label{ME1} 
\ee
 then $\hat\bF=\d\hat\bA$, $\hat\bA=\frac{\phi}{\lambda}\d u$ 
	 with
	\be 
	\hat F_{iu}=\left(\frac{\H}{\lambda}\right),_i=\frac{\lambda\H,_i-\lambda,_i \H}{\lambda^2}
	 \label{Fiu_boosted}
	\ee
	satisfies the Maxwell field equations iff 
		\be
	\left(\frac{\phi}{\lambda}\right),_{ii} 
		=0 .\label{ME2}
	\ee
	Using~\eqref{ME1}, eq.~\eqref{ME2} can be rewritten as 
	\be 
	\left(\frac{\lambda,_i\phi^2}{\lambda^2}\right),_i=0 .
	\label{nonuniq_1b}
	\ee 
 Provided $\lambda$ satisfies~\eqref{nonuniq_1b},  the boost produces a non-equivalent single copy  $\hat\bF=\d\hat\bA$  (w.r.t. $\hat\bl=\lambda\d u$) of the same KS metric. In the special case $\lambda=\lambda(u)$,\footnote{This corresponds to a boost with $D\lambda=0=\delta_i\lambda$, which preserves the gauge condition $L_{1i}=0$ assumed in sections~\ref{subsec_KundtNP_gen} and \ref{subsubsec_gauge_pp} (cf.~\eqref{boost_coef}).} eq.~\eqref{Fiu_boosted} reduces to $\hat F_{iu}=\frac{F_{iu}}{\lambda}=\frac{\H,_i}{\lambda}$, and $F^{ab}_{\phantom{ab};b}=0$ implies $\hat F^{ab}_{\phantom{ab};b}=0$ (see \cite{Urbantke79} for related comments in four dimensions). Thus in this case, 
	two null electromagnetic fields related by a rescaling by any function $\lambda(u)$ can represent two different single copies of the same gravitational field.
	  For example, when the $u$-dependence of $\H$ is factorized, this freedom can be used to make $\hat\bF$ $u$-independent and thus clearly physically distinct from $\bF$.

From a complementary viewpoint, assume $\H_1 $ and $\H_2$ are harmonic functions. Let $\bA=\H_1 \bl=\H_1 \d u$ be a single copy of the metric~\eqref{eqPP_KS} (with $\H^{(0)}=\H_1$). Choose $\lambda = \frac{\H_1}{\H_2}$. Then $\hat \bA = \hat\H\hat \bl=\frac{\H_1}{\lambda^2}\hat\bl =\frac{{(\H_2})^2}{\H_1} \hat \bl=\H_2 \d u$ (cf.~\eqref{rescal1}) is also a single copy corresponding to the same KS metric (see~\eqref{ME2} with $\phi\to\phi_1$). Since the previous observation applies to {\em any} solutions of the Laplace equation (except at spacetime points where $\H_2=0$), this also means that any type~N \pp wave spacetime \eqref{eqPP_KS} corresponds via  a double copy to any aligned null electromagnetic field and vice versa. 

The same argument as above can be used to show an infinite non-uniqueness of the single copy also for Kundt waves, now using \eqref{rescal1}, \eqref{case1_A} with $\lambda=\frac{\psi_1}{\psi_2}$ and (cf.~\eqref{kundtw_db}) 
\be
	\d s^2 =\d s^2_{\mbox{\tiny flat}}  -2 \y\ff_1\d u^2
\ee
with $\triangle \ff_1=0=\triangle \ff_2$ ($\d s^2_{\mbox{\tiny flat}}$ is given by~\eqref{VSI} with \eqref{VSI_K_W2}--\eqref{VSI_K_H}, \eqref{H0_Kundt}).

\subsection{Case 2. -- non-null vs. null-field single copy}
	
\label{sec_nonuni2}

Consider now a \pp wave metric~\eqref{VSIpp}. By a coordinate transformation \cite{Coleyetal06} 
\be
\hat u=g(u),\ \ \hat v=\frac{v}{g,_u},\ \ \hat x^i=x^i , 
\label{hat_u}
\ee
one retains the metric form~\eqref{VSIpp} except for adding a linear term in $v$ to $H^{(0)}$
\be
\hat H=\hat H^{(1)} \hat v + \hat H^{(0)}, \qquad
\hat H^{(1)}(u) =\frac{g,_{uu}}{g,_u^2}, \qquad  \hat H^{(0)}=\frac{H^{(0)}}{g,_u^2}. 
\label{hat_H}
\ee

The metric~{\eqref{VSIpp} for $H=H^{(0)}_{\mbox{\tiny flat}}+h(u)v$ is flat.\footnote{From now on let us omit hats on transformed quantities to keep the notation simple.}
Thus the term $h(u)  v$ can be either added to the flat part of the metric \eqref{KS} (which leads to a usual null electromagnetic plane-wave single copy), see section \ref{subsec_pp_gauge}, or kept  in the KS part of the metric (thus $\H^{(1)}=-H^{(1)}$),
see section \ref{sub_pp_nonnull}, 
\be 
	\d s^2= 
2\d u 
\d v 
+ \delta_{ij} \d x^i\d x^j 
	- 2 (v\H^{(1)}+\H^{(0)})\d u^2.  \label{g_nonnull}
\ee 
where $\H^{(0)}=\H^{(0)}(u,x^i)$. Here $\bl=\d u$ and $\ell_{a;b}=-\H^{(1)}\ell_a\ell_b$.  

While the term $\H^{(1)}$ in~\eqref{g_nonnull} does not affect the spacetime curvature, it does produce also a boost-weight zero component of $\bF$
 (cf. \eqref{elmgF})
\be 
 \bA=( v \H^{(1)}+\H^{(0)})\d  u, \qquad   F_{01} = \H^{(1)},\ \  F_{1i}=-\delta_i\H ,
 \label{hat_A}
\ee 
implying that  $\bF$ is  non-null, as can be also seen from the non-vanishing invariant $F_{ab} F^{ab}=-2(\H^{(1)})^2$.  However, note that double-copy compatibility conditions imply $\H^{(1)}=$const, see section  \ref{sub_pp_nonnull}.
Then the Einstein and Maxwell equations reduce to 
	\be
	\H^{(0)},_{ii}=0 .
	\ee

\section*{Acknowledgments}

This work has been supported by research plan RVO: 67985840.

\renewcommand{\thesection}{\Alph{section}}
\setcounter{section}{0}

\renewcommand{\theequation}{{\thesection}\arabic{equation}}

\section{Formulas of the higher-dimensional Newman-Penrose formalism employed in this paper }
\setcounter{equation}{0}

\label{app_NP}

Here, let us complement  section~\ref{sec_prelim} by giving a concise summary of the higher-dimensional Newman-Penrose formalism as needed throughout the paper.

Under the boost transformation~\eqref{boosts}, the Ricci rotation coefficients transform as \cite{OrtPraPra07}
\beqn
& & \hspace{-2.3cm} \hL_{11}=\lambda^{-1}L_{11}+\lambda^{-2}\Delta\lambda , \quad \hL_{10}=\lambda L_{10}+D\lambda , \quad \hL_{1i}=L_{1i}+\lambda^{-1}\delta_i\lambda , \quad \hL_{i1}=L_{i1} , \nonumber \label{boosts2} \\ 
& & \hspace{-2.3cm} \hL_{i0}=\lambda^2L_{i0} , \quad \hL_{ij}=\lambda L_{ij} , \qquad \hN_{i1}=\lambda^{-2}N_{i1} , \quad \hN_{i0}=N_{i0} , \quad   \hN_{ij}=\lambda^{-1}N_{ij} ,
\label{boost_coef}\\
& & \hspace{-2.3cm} \hM{i}{j}{1}=\lambda^{-1}\M{i}{j}{1} , \qquad \hM{i}{j}{0}=\lambda\M{i}{j}{0} ,\qquad \hM{i}{j}{k}=\M{i}{j}{k} . \nonumber 
\eeqn

In addition to boosts~\eqref{boosts}, the remaining Lorentz transformations can be described in terms of null rotations about $\bl$
\be
 \hbl=\bl, \qquad \hbn =\bn+z_i\bm^{(i)} -\frac{1}{2} z^2\bl , \qquad \hbm{i} =\bm^{(i)} -z_i\bl ,
    \label{nullrot}
\ee
where the $z_i$ are real functions and $z^2\equiv z_iz^i$, null rotations about $\bn$ (obtained by simply interchanging $\bl\leftrightarrow\bn$ in~\eqref{nullrot}) and spins
\be
 \hbl =  \bl, \qquad \hbn = \bn, \qquad \hbm{i} =  X^{i}_{\ j} \bm^{(j)} , 
 \label{spins}
\ee
where the $X^{i}_{\ j}$ are $(n-2)\times(n-2)$ orthogonal matrices. For brevity, we just refer to \cite{OrtPraPra07} for the corresponding transformation formulae of the Ricci rotation coefficients.

Commutators of operators \eqref{covder} read \cite{Coleyetal04vsi} 
\bea
\Delta D - D \Delta &=& L_{11} D + L_{i1} \delta_i ,\label{com_TD} \\
\delta_i D - D \delta_i &=& L_{1i} D  + L_{ji} \delta_j, 
\label{com_Tdi}\\
\delta_i \Delta - \Delta \delta_i  &=& N_{i1} D + (L_{i1}-L_{1i}) \Delta + (N_{ji}-\MM{i}{j1}) \delta_j, \\
\delta_i \delta_j - \delta_j \delta_i &=& (N_{ij}-N_{ji}) D + (L_{ij}-L_{ji})\Delta
+ (\MM{j}{ki}-\MM{i}{kj})\delta_k.\label{com_dd}
\eea

From now on, we restrict ourselves to Weyl type N, Ricci-flat Kundt spacetimes and use an affine parametrization and a parallelly propagated frame (cf.~\eqref{pt}, \eqref{kundt}). 
The Ricci identity  gives \cite{OrtPraPra07} (cf. also, e.g., appendix~A of \cite{OrtPra16})
\beqn
& & DL_{1i}=0=DL_{i1} , \qquad \delta_jL_{i1}=L_{i1}L_{j1}-L_{k1}\overset{k}{M}_{ij}, \qquad \delta_{[j|} L_{1|i]} =  -  L_{1k} \MM{k}{[ij]} , \label{Ricci_dL} \\
& & D\overset{i}{M}_{jk}=0 , \qquad DL_{11}=-L_{1i}L_{i1} , \qquad DN_{ij}=0 , \label{Ricci_DN}\\
& & \bigtriangleup L_{1i} - \delta_i L_{11} =   
 L_{11}(L_{1i}- L_{i1}) -  L_{j1} N_{ji} -L_{1j} (N_{ji}+\M{j}{i}{1}) .\label{Ricci_deltaL11}
 \eeqn

It is also useful to note that \cite{Pravdaetal04}
\be
	\ell_{a ; b }=L_{11} \ell_a \ell_b +L_{1i} \ell_a \msub{i}{b}+L_{i1} \msub{i}{a} \ell_b   \qquad (L_{ij}=0) .
	\label{dl_kundt} 
\ee

\section{Curvature for KS-Kundt spacetimes }
\setcounter{equation}{0}

\label{app_Riem}

Here we consider KS spacetimes~\eqref{KS}, in which $\bl$ is Kundt (and thus geodesic) and the background geometry $\bet$ is flat, but without imposing the Einstein equations. We take an affine parameter and a frame $\{\bl,\bn,\bm_{(i)}\}$ parallelly transported along $\bl$ (cf.~\eqref{affine}, \eqref{pt}). The derivative operators~\eqref{covder} in the full and the flat geometries are related by \cite{OrtPraPra09}
\be
 D=\tilde D , \qquad \delta_i=\tilde \delta_i  , \qquad \Delta=\tilde \Delta+\H\tilde D ,
 \label{covder_KS}
\ee
with a corresponding relation for the frame vectors (in particular $n^a=\tilde n^a+\H\ell^a$ and thus $n_a=\tilde n_a-\H\ell_a$). In the special case of a scalar function $f$ which satisfies $Df=0$, one thus has $\Delta f=\tilde\Delta f$.
 
The (possibly) non-zero Ricci rotation coefficients~\eqref{Ricci_rot} for a KS-Kundt geometry~\eqref{KS}, \eqref{kundt} read
\beqn
 & & \M{i}{j}{k}=\Mt{i}{j}{k} \label{Li0_KS} , \qquad L_{i1}=\tilde L_{i1} , \qquad L_{1i}=\tilde L_{1i} , \qquad N_{ij}=\tilde N_{ij} , \label{Nij_KS} \\
 & & \M{i}{j}{1}=\Mt{i}{j}{1} , \qquad  L_{11}=\tilde L_{11}-\tilde D \H , \qquad N_{i1}=\tilde N_{i1}+\H\big(2\tilde L_{1i}-\tilde L_{i1}\big)+\tilde \delta_i\H ,  \label{L11_KS} 
\eeqn 
where tilded quantities refer to the flat background geometry $\bet$ (defined by $\H=0$) evaluated in the tilded frame.

Thanks to~\eqref{kundt}, the only non-zero components of the Riemann tensor are \cite{OrtPraPra09}
\beqn
  & & R_{0101} = D^2\H , \label{R0101_KS} \\  
  & & R_{011i} = \big(-\delta_i D  + 2 L_{[i1]}D\big) \H  , \label{R011i_KS} \\
	& &  R_{1i1j}=\big[\delta_{(i}\delta_{j)}+\M{k}{(i}{j)}\delta_k +2(2L_{1(j|}-L_{(j|1})\delta_{|i)}+N_{(ij)}D\big]\H  \nonumber \\
  & & \qquad\qquad {}+2 \H\big(\delta_{(i|}L_{1|j)}-2L_{1(i}L_{j)1}+2L_{1i}L_{1j}+L_{1k}\M{k}{(i}{j)}\big) . \label{R1i1j_KS} 
\eeqn

It follows that the Ricci tensor is given by \cite{OrtPraPra09}
\beqn
 R_{01} & = & -D^2\H , \label{R01} \\  
 R_{1i} & = & \big(-\delta_iD +2L_{[i1]}D\big) \H , \label{R1i} \\  
 R_{11} & = & \big[\delta_i\delta_i +N_{ii}D +(4L_{1j}-2L_{j1}-\M{i}{j}{i})\delta_j\big] \H \nonumber \\
				& & {}+2 \H\big[2\delta_iL_{[1i]}+4L_{1i}L_{[1i]}+L_{i1}L_{i1} +2L_{[1j]}\M{j}{i}{i}\big] . \label{R11} 
\eeqn
(The Ricci identity \cite{OrtPraPra07} has also been used in~\eqref{R11}.) We observe that the Riemann and Ricci tensors are linear in $\H$ (see \cite{GurGur75,DerGur86} for a similar observation for the mixed coordinate components $R^a_{\phantom{a}b}$ of general KS spacetimes).

In passing, let us observe that, for vacuum solutions (eqs.~\eqref{H_K}, \eqref{E1i_K2}, \eqref{E11_K2}) in the gauge $L_{1i}=0$ (cf. section~\ref{subsubsec_gauge}), the only non-zero component~\eqref{R1i1j_KS} of the curvature tensor reduces to
\be
	C_{1i1j}=R_{1i1j}=\big(\delta_{(i}\delta_{j)}+\M{k}{(i}{j)}\delta_k-2L_{(j|1}\delta_{|i)}\big)\H^{(0)}+N_{(ij)}\H^{(1)} \qquad (L_{1i}=0) . \label{R1i1j_gauge} 
\ee

\section{Debney coordinates for all KS-Kundt metrics, and double copy}

\label{app_debney}

Here we show how to extend to arbitrary dimension the $n=4$ coordinates and frame of \cite{Debney73,Debney74}, in connection to the general analysis of sections~\ref{sec_KundtNP} and \ref{sec_pp}.

\subsection{Cartesian coordinates and adapted frame}

In Cartesian coordinates, the KS line-element~\eqref{KS} can we written as (cf.~\cite{DebKerSch69,Debney73,Debney74} for $n=4$)
\be
 \d s^2=2\d\tu\d\tilde r+ \delta_{ij}\d x^i\d x^j -2\H\bl\otimes\bl ,
 \label{KS_debney}
\ee
where 
\be
 \ell_a\d x^a=\d\tu+Y_i\d x^i-\frac{1}{2}Y^2\d\tilde r \qquad (Y^2\equiv Y_jY^j) ,
 \label{l_debney}
\ee
represents the most general (up to a boost~\eqref{boosts}) null vector field, and $\H$, $Y^i$ are spacetime functions (for notational convenience we identify $Y^i=Y_i$,  and similarly below for the functions $X^i=X_i$ defined in~\eqref{Xi}).

Along with $\bl$, a null frame (as defined in section~\ref{sec_prelim}) can be completed by taking 
\be
  n_a\d x^a=\d\tilde  r-\H\ell_a\d x^a , \qquad m^{(i)}_a\d x^a=\d x^i-Y^i\d\tilde  r.
	\label{n_debney}
\ee 

The derivative operators then read 
\be
 D=\pa_{\tilde r}+Y^i\pa_{x^i}-\frac{1}{2}Y^2\pa_{\tu} , \qquad \Delta=\pa_{\tu}+\H D , \qquad \delta_i=\pa_{x^i}-Y_i\pa_{\tu} .
\ee
For later use, let us observe that $\delta_i\tilde r=0$. 

In the above frame one finds (cf.~\cite{Debney73,Debney74} in four dimensions and~\cite{Krssak09} in any dimensions\footnote{There is a typo in (3.43,\cite{Krssak09}).}) 
\beqn
 & & L_{i0}=DY_i , \qquad L_{10}=0 , \qquad L_{ij}=\delta_j Y_i , \qquad L_{i1}=(\Delta-\H D)Y_i=Y_{i,\tu} , \quad L_{1i}=-\H L_{i0} , \\
 & & L_{11}=-D\H , \qquad \MM{i}{jk}=0 , \qquad \MM{i}{j0}=0 , \qquad N_{i0}=0 ,\qquad N_{ij}=\H L_{ji} , \\
 & & \MM{i}{j1}=2\H L_{[ij]} , \qquad N_{i1}=\delta_i\H-\H\Delta Y_i .
\eeqn 

Requiring $\bl$ to be Kundt means (cf.~\eqref{pt}, \eqref{kundt}) $DY_i=0=\delta_j Y_i$, such that
\be
 \d Y_i=(\Delta Y_i)\ell_a\d x^a .
 \label{dY}
\ee
The above Ricci rotation coefficients then simplify to
\beqn
 & & L_{i0}=0 , \qquad L_{10}=0 , \qquad L_{ij}=0 , \qquad L_{i1}=\Delta Y_i , \qquad L_{1i}=0 , \label{Li0_deb} \\
 & & L_{11}=-D\H , \qquad \MM{i}{jk}=0 ,  \qquad \MM{i}{j0}=0 , \qquad N_{i0}=0  , \qquad N_{ij}=0 , \label{L11_deb} \\
 & & \MM{i}{j1}=0 , \qquad N_{i1}=\delta_i\H-\H\Delta Y_i .
\eeqn 
We note that, by construction, the frame in use is parallelly transported and $\bl$ is affinely parametrized.

The Riemann~\eqref{R0101_KS}--\eqref{R1i1j_KS} and Ricci~\eqref{R01}--\eqref{R11} tensors then reduce to
\beqn
  & & R_{0101} = D^2\H , \qquad R_{011i} = \big(-\delta_i D  + L_{i1}D\big) \H  ,  \qquad R_{1i1j}=\big[\delta_{(i}\delta_{j)}-2L_{(j|1})\delta_{|i)}\big]\H , \label{R1i1j_KS2} \\
	& & R_{01}=-D^2\H , \qquad R_{1i}=\big(-\delta_iD +L_{i1}D\big) \H , \qquad R_{11}=\big[\delta_i\delta_i-2L_{i1}\delta_i\big]\H  , \label{R11_2} 
\eeqn
where in $R_{11}$ we used the second of~\eqref{Ricci_dL}. So far we have not imposed any field equations and the KS-Kundt metric~\eqref{KS_debney} (with~\eqref{dY}) is, at this stage, of Weyl (and Riemann) type~II.

\subsection{Debney coordinates, vacuum solutions and double copy}

Defining new coordinates $(u,r=\tilde r,y^i)$ (cf.~\cite{Debney73,Debney74}) 
\beqn
 u\equiv \tu+Y_ix^i-\frac{1}{2}Y^2  \tilde r , \qquad y^i\equiv x^i- \tilde rY^i , 
\eeqn
the basis of 1-forms~\eqref{l_debney}, \eqref{n_debney} takes the form
\beqn
 & & \ell_a\d x^a=\frac{\d u}{1+X_iy^i} , \qquad   n_a\d x^a=\d r-\H\ell_a\d x^a , \qquad m^{(i)}_a\d x^a=\d y^i+rX^i\ell_a\d x^a , \label{coframe_deb} \\
 & & X_i\equiv\Delta Y_i=L_{i1} , \label{Xi}
\eeqn 
and the derivative operators read simply 
\be
 D=\pa_r , \qquad \Delta=\big(1+X_iy^i\big)\pa_u-rX^i\pa_{y^i}+\H D , \qquad \delta_i=\pa_{y^i} .
\ee
In particular, $r$ is the same coordinate as defined in~\eqref{affine}. Note that in these coordinates $Y_i=Y_i(u)$, $\d Y_i=X_i\ell_a\d x^a$ and thus $X_i=\big(1+X_jy^j\big)Y_{i,u}$, which also gives $1+X_jy^j=\big(1-Y_{j,u}y^j\big)^{-1}$. We further have $\delta_ir=0$, as in the set-up of section~\ref{subsubsec_gauge}. There remains a freedom of boosts~\eqref{boosts} with $\lambda=\lambda(u)$ and spins~\eqref{spins} with $X^i_{\phantom{i}j}=X^i_{\phantom{i}j}(u)$, which both preserve \eqref{Li0_deb}, \eqref{L11_deb}, except for $\hL_{11}=\lambda^{-1}L_{11}+\lambda^{-2}\Delta\lambda$ (cf.~\eqref{boosts2}).  Since $L_{i1}=X_i$ is invariant under such transformations, $\bl$ can be (parallel to) a covariantly constant null vector field iff $X_i=0\Leftrightarrow Y_{i,u}=0$.

By~\eqref{R11_2}, in vacuum one obtains
\be
	\H=  r\H^{(1)}+\H^{(0)} , \label{E1_deb}
\ee	
with
\beqn
	& & \H^{(1)}=\frac{f(u)}{1-Y_{j,u}y^j} , \label{E2_deb} \\
  & & \big[(1-Y_{j,u}y^j)\H^{(0)}\big]_{,y^iy^i}=0 , \qquad \H^{(0)}_{,r}=0 .	 \label{E3_deb} 
\eeqn	
where $f(u)$ is an integration function. Recall that the first of~\eqref{E3_deb} is equivalent to the wave equation for $\H^{(0)}$ in flat space (section~\ref{sec_prelim}).

The only non-zero components of the curvature tensor become 
\be
 C_{1i1j}=R_{1i1j}=\big[\delta_{(i}\delta_{j)}-2X_{(j}\delta_{i)}\big]\H^{(0)} ,
 \label{C1i1j}
\ee
such that the Weyl type is N (as expected from the general results mentioned in section~\ref{intro}).

Since the function $\H^{(1)}$ does not enter the curvature~\eqref{C1i1j}, the most general vacuum KS-Kundt line element~\eqref{KS_debney} can be rewritten as
\beqn
 & & \d s^2=\d s^2_{\mbox{\tiny flat}}-2\H^{(0)}\bl\otimes\bl , \qquad \bl=\big(1-Y_{j,u}y^j\big)\d u , \label{KS_debney2} \\
 & & \d s^2_{\mbox{\tiny flat}}=2\big(1-Y_{j,u}y^j\big)\d u\d r+\big(\d y^i+rY^i_{\phantom{i},u}\d u\big)\big(\d y^i+rY^i_{\phantom{i},u}\d u\big)-2r\H^{(1)}\bl\otimes\bl , \label{KS_debney2_back}
\eeqn
where $\H^{(1)}$ and $\H^{(0)}$ obey \eqref{E2_deb}, \eqref{E3_deb}. The function $\H^{(1)}$ has been effectively absorbed into the background part of the metric and one can thus apply the (null field) double copy as described in section~\ref{subsec_KundtNP_gen} for $X_i\neq0$ and in section~\ref{subsec_pp_gauge} for $X_i=0$.
As mentioned above, recall that $\bl$ can be boosted with an arbitrary $\lambda(u)$, which corresponds to the case~1. of the non-uniqueness discussed in section~\ref{nonuni1} (see also section~\ref{subsec_KundtNP_gen}).

The case $X_i=0$ corresponds to vacuum \pp waves (i.e., $Y_{i,u}=0$, cf. above), for which~\eqref{E2_deb}, \eqref{E3_deb} reduce to
\be
	  \H^{(1)}=f(u) , \qquad \H^{(0)}_{,y^iy^i}=0 . 
		\label{debn_pp}
\ee 
For certain applications, it is useful to recall that $f(u)$ can be rescaled to an arbitrary constant (including zero) with a transformation of the type~\eqref{hat_u}.
Here $\bl=\d u$ and $L_{11}=-f(u)$, which can be set to zero by a suitable boost, thus obtaining a covariantly constant $\hat\bl$.  It is worth recalling that for \pp waves the double copy can additionally be applied also to the alternative form of the metric (with~\eqref{debn_pp} and $f=$const)
\beqn
 & & \d s^2=\d s^2_{\mbox{\tiny flat}}-2\big(r\H^{(1)}+\H^{(0)}\big)\bl\otimes\bl , \qquad \bl=\d u , \label{pp_deb1} \\
 & & \d s^2_{\mbox{\tiny flat}}=2\d u\d r+\d y^i\d y^i , \label{pp_deb2} 
\eeqn
as shown in section~\ref{sub_pp_nonnull}. This gives rise to a non-null gauge field and corresponds to the case~2. of the non-uniqueness discussed in section~\ref{sec_nonuni2}.

\subsection{Relation to Kundt coordinates}

After a null rotation~\eqref{nullrot} and a boost~\eqref{boosts} of the coframe~\eqref{coframe_deb} with parameters
\be
  z_i=rP^{-1}Y_{i,u} , \qquad \lambda=P^{-1} , \qquad P\equiv 1-Y_{j,u}y^j ,
\ee 	
accompanied by a coordinate transformation 
\be
 r=\frac{v}{P} , 
\ee
one arrives at the following canonical Kundt coframe
\be
  \hat\ell_a\d x^a=\d u , \qquad   \hat n_a\d x^a=\d v+\frac{2v}{P}Y_{j,u}\d y^j+\left(\frac{v^2}{2P^2}Y^i_{\phantom{i},u}Y^i_{\phantom{i},u}+\frac{v}{P}Y_{j,uu}y^j-P^2\H\right)\d u , \qquad \hat m^{(i)}_a\d x^a=\d y^i  , 
\ee 
i.e., one obtains the standard Kundt metric~\eqref{VSI} with
\be
  W_i=\frac{2v}{P}Y_{i,u} , \qquad H=\frac{v^2}{2P^2}Y^i_{\phantom{i},u}Y^i_{\phantom{i},u}+\frac{v}{P}Y_{j,uu}y^j-P^2\H  .
	\label{W_H_deb}
\ee 
The above result covers all KS-Kundt geometries. In vacuum, one needs to impose~\eqref{E1_deb}, which now reads
\be
	\H=\frac{v}{P}\H^{(1)}+\H^{(0)} , 
	\label{phi_deb}
\ee	
with \eqref{E2_deb}, \eqref{E3_deb}. This choice of $\H$ also implies the Weyl type~N.

To conclude, let us present the relation between the general (type~N) Kundt coordinates constructed above and the particular canonical forms considered separately for Kundt and \pp waves in sections~\ref{Kundt_coords} and \ref{subsec_pp}.

\subsubsection{Kundt waves ($Y_{i,u}\neq0$)} 

The canonical form of \eqref{VSI_K_W2}--\eqref{VSI_K_H} of \cite{Coleyetal06} can be obtained from \eqref{W_H_deb}, \eqref{phi_deb} (with \eqref{E2_deb}, \eqref{E3_deb}) by performing a spatial rotation $y^i=\R^i_{\phantom{i}j}(u)\tilde y^j$ such that $\R^i_{\phantom{i}j}Y_{i,u}=A(u)\delta_{2,j}$ ($\R^i_{\phantom{i}j}$ is an orthogonal matrix and $A^2=Y_{i,u}Y_{i,u}$), followed by the redefinitions $\tilde y^2=x+A^{-1}$ (such that $P=-Ax$), $v\mapsto v+g(u,x,\tilde y^m)$ (to get rid of a $v$-independent term in $W_2$; $m=3,\ldots,n-1$) and finally a transformation of the form $u\mapsto h(u)$, $v\mapsto v/\dot h$ (to get rid of a term proportional to $v$ in $H$)
\cite{Stephanibook,Coleyetal06}.

\subsubsection{\pp waves ($Y_{i,u}=0$)}

As mentioned above, \pp waves correspond to $Y_{i,u}=0$ and thus $P=1$ and $r=v$, for which \eqref{W_H_deb}, \eqref{phi_deb}, \eqref{E2_deb}, \eqref{E3_deb} give
\be
  W_i=0 , \qquad H=-vf(u)-\H^{(0)} ,
\ee 
which is equivalent to~\eqref{pp_deb1}, \eqref{pp_deb2}. However, a transformation $u\mapsto h(u)$, $v\mapsto v/\dot h$ \cite{Stephanibook,Coleyetal06} (cf.~\eqref{hat_u}, \eqref{hat_H}) can be used to arrive at the canonical metric functions~\eqref{pp_W}, \eqref{pp_H} of \cite{Schimming74} (see also \cite{Horowitz90,Sokolowski91,KucOrt19}) with $H^{(0)}=-\H^{(0)}$ (after which a boost $\bl\mapsto\bl/\dot h$ gives rise to a covariantly constant null vector field).

\section{Double copy in modified theories: an example}

\label{app_example}

Here we illustrate the ``universal'' character of the KS-Kundt double copy mentioned in section~\ref{intro} using a specific modified theory as an example. On the gravity side, let us consider general relativity with an additional Gauss-Bonnet term, described by the action
\be
S_G=\int\d^n x\sqrt{-g}\left[\frac{1}{\kappa}(R-2\Lambda)+\gamma I_{GB}\right] , \qquad I_{GB}\equiv R_{abcd}R^{abcd}-4R_{ab}R^{ab}+R^2 ,
\label{GB}
\ee
where $\kappa$ and $\gamma$ are a constant parameters. This gives rise to the gravity field equations
\be
\frac{1}{\kappa}\left(R_{ab}-\frac{1}{2}Rg_{ab}+\Lambda g_{ab}\right)+2\gamma\left(RR_{ab}-2R_{acbd}R^{cd}+R_{acde}R_b^{\ cde}-2R_{ac}R_b^{\ c}-\frac{1}{4}I_{GB}g_{ab}\right)=0 .
\label{E_GB}
\ee
Since the Kundt waves~\eqref{VSI} (with \eqref{VSI_K_W2}--\eqref{eqEinstein_Kundt}) and the \pp waves~\eqref{VSIpp}, \eqref{eqEinstein_PP} are Ricci-flat spacetimes of Weyl type~N, the Einstein and the Gauss-Bonnet terms of~\eqref{E_GB} vanish separately, which implies that both such metrics are vacuum solutions of the theory~\eqref{GB} \cite{BouDes85,GleDot05,PraPra08} (see also \cite{HerPraPra14} for more general results).

As for the electromagnetic theory, let us modify the standard Maxwellian action by an additional ``$F^4$'' term (cf., e.g., \cite{CveNojOdi02,MaeHasMar10} and references therein), namely 
\be
S_M=\int\d^n x\sqrt{-g}\left[-\frac{1}{4}F_{ab}F^{ab}+c_1(F_{ab}F^{ab})^2+c_2F_{ab}F^{bc}F_{cd}F^{da}\right] ,
\label{F4} 
\ee
where $c_1$ and $c_2$ are constants. This theory has non-linear equations of motion of the form
\be
\nabla_b\left[-F^{ab}+8c_1(F_{cd}F^{cd})F^{ab}+8c_2F^{ac}F_{cd}F^{db}\right]=0 .
\ee
The 2-form $\bF=\d\bA$ with~\eqref{potA} obeying~\eqref{Maxwell} is a null field solution of Maxwell's equations, which implies that $\nabla_b F^{ab}=0$, $F_{cd}F^{cd}=0$ and $F^{ac}F_{cd}F^{db}=0$ separately (for the latter conclusion see \cite{OrtPra16} and Proposition~2.4 of \cite{HerOrtPra18}). Therefore $\bF$ is also a solution of the modified theory~\eqref{F4} (see also \cite{OrtPra16,OrtPra18,HerOrtPra18} for more general results). The fact that such $\bF$ is a single copy of the metrics \eqref{VSI} (with \eqref{VSI_K_W2}--\eqref{eqEinstein_Kundt}) and~\eqref{VSIpp}, \eqref{eqEinstein_PP} follows as in sections~\ref{sec_Kundtline} and \ref{subsec_pp}.

Note that, in the spirit of the KS double copy \cite{MonOCoWhi14}, the electromagnetic field has been treated as a test field above. However, the same arguments can be extended to include back-reaction, i.e., the coupled theory $S_G+S_M$ (cf.~Theorem~3.3 of \cite{KucOrt19}).

\providecommand{\href}[2]{#2}\begingroup\raggedright\endgroup

\end{document}